\documentclass[journal,twoside,web]{ieeecolor}
\usepackage{generic}
\usepackage{cite}
\usepackage{amsmath,amssymb,amsfonts}
\usepackage{algorithmic}
\usepackage{graphicx}
\usepackage{textcomp}
\usepackage{booktabs}
\usepackage{subcaption}
\usepackage{multirow}
\usepackage{arydshln}

\usepackage[outdir=./]{epstopdf}

\usepackage[labelformat=simple]{subcaption}    

\def\BibTeX{
{\rm B\kern-.05em
{\sc i\kern-.025em b}
\kern-.08em T\kern-.1667em\lower.7ex\hbox{E}\kern-.125emX}
}
\begin{document}
\title{CoRe-Sleep: A Multimodal Fusion Framework for Time Series Robust to Imperfect Modalities.}
\author{Konstantinos Kontras, Christos Chatzichristos, Huy Phan, \\ Johan Suykens and Maarten De Vos
\thanks{
This project has received funding from the Flemish Government (AI Research Program) and from the FWO ( 'Artificial Intelligence (AI) for data-driven personalised medicine', G0C9623N and 'Deep, personalized epileptic seizure detection', G0D8321N ) and Leuven.AI Institute.}
\thanks{K. Kontras, C. Chatzichristos, J. Suykens and M. De Vos are with the Stadius group in the Department of Electrical Engineering at KU Leuven University, 3000 Leuven, Belgium. M. De Vos is also with the Department of Development and Regeneration, KU Leuven University (e-mail: konstantinos.kontras@kuleuven.be; christos.chatzichristos@kuleuven.be; johan.suykens@kuleuven.be; maarten.devos@kuleuven.be). }
\thanks{H. Phan is with Amazon Alexa, Cambridge, USA (e-mail: huypq@amazon.co.uk).}}

\maketitle

\begin{abstract}
Sleep abnormalities can have severe health consequences. Automated sleep staging, i.e. labelling the sequence of sleep stages from the patient's physiological recordings, could simplify the diagnostic process. Previous work on automated sleep staging has achieved great results, mainly relying on the EEG signal. However, often multiple sources of information are available beyond EEG. This can be particularly beneficial when the EEG recordings are noisy or even missing completely. In this paper, we propose CoRe-Sleep, a Coordinated Representation multimodal fusion network that is particularly focused on improving the robustness of signal analysis on imperfect data. We demonstrate how appropriately handling multimodal information can be the key to achieving such robustness. CoRe-Sleep tolerates noisy or missing modalities segments, allowing training on incomplete data. Additionally, it shows state-of-the-art performance when testing on both multimodal and unimodal data using a single model on SHHS-1, the largest publicly available study that includes sleep stage labels. The results indicate that training the model on multimodal data does positively influence performance when tested on unimodal data. This work aims at bridging the gap between automated analysis tools and their clinical utility. 
\end{abstract}


\section{Introduction}
\label{sec:introduction}
\IEEEPARstart{S}{leep} is vital for our well-being. It has been linked to various brain and mental health diseases \cite{brzecka2018sleepalzheimer}, particularly in the elderly population \cite{patel2022physiology}. Sleep studies typically involve recording patients' sleep using various sensors in a clinical setting. These sensors measure various physiological signals, such as Electroencephalogram (EEG), Electrocardiogram (ECG), Electrooculogram (EOG), Electromyogram (EMG) and respiration.

During the analysis of the patient's sleep, a sleep stage label is given to every 30 seconds for the duration of the recording. The American Academy of Sleep Medicine (AASM) scoring standard \cite{berry2012aasm} recommends assigning the following labels for adult sleep: Wake, Rapid Eye Movement (REM) and the three sleep phases N1, N2, and N3 related to lighter and deeper sleep. Extracting such a sequence of sleep labels (which is called the hypnogram) is time-consuming for experts, hence research focuses on automating it.   

Previous research on automating sleep staging has primarily focused on using a single modality, EEG. Initial approaches used handcrafted features \cite{alickovic2018ensemble, li2012electroencephalogram, bajaj2013automatic, fraiwan2012automated}, based on prior knowledge about the characteristics of EEG. Recent methods employ Neural Networks (NNs) to automatically extract these features, for example with Convolutional NNs (CNN) \cite{tsinalis2016automatic, tinysleepnet, supratak2017deepsleepnet}, Recurrent NNs (RNN) \cite{phan2019seqsleepnet, phan2021xsleepnet}, Transformers (TF) \cite{phan2022sleeptransformer, supratak2017deepsleepnet, lee2022sleepyco, pradeepkumar2022towards, eldele2021attention} and more \cite{perslev2021u,jia2020graphsleepnet}. 

Efforts have been made to exploit the use of multimodal input for sleep staging. Multimodal data, obtained from different sources, offer interrelated observations of the underlying phenomenon. They potentially contain complementary information for the identification of the underlying data distribution \cite{ramachandram2017deep, lahat2015multimodal}. Concatenating EEG, EOG, and EMG signals as model input, instead of EEG alone, has been shown to slightly improve performance in previous studies \cite{phan2019seqsleepnet, phan2021xsleepnet}. However, there has been limited research on how to optimize the available multimodal input. It remains crucial how different combining methods (fusion approaches) could improve results.

Multimodal information can be beneficial for sleep staging. While exploiting multimodal input might only marginally outperform the best single modality (i.e. the EEG), it might enhance the robustness of the system when this single modality fails. For example, when EEG is noisy or missing, multimodal processing can lead to more accurate results than relying on EEG alone. 

Missing and noisy data have been handled with imputation and denoising strategies respectively. Imputing missing values has been widely explored in statistics either for a few missing values \cite{buck1960method, rubin1978multiple} or for higher dimensional data \cite{white2011multiple, stekhoven2012missforest, bischke2018overcoming, gans_eeg_imputation, comas2020learning, cao2018brits, yoon2018estimating}. Recently, imputation has been achieved by masking the missing data with a learnable embedding \cite{rnn_noimputation, shen2019brain}, allowing the network to interpolate the values based on the remaining input. In the case of noisy samples, data are retained but there might be ambiguity about the amount of useful information they contain. Previous approaches facing noisy EEG include either an explicit \cite{nolan2010faster, bigdely2015prep, jas2017autoreject} EEG-tailored denoising pre-phase or a more generic implicit one \cite{banville2022robust}. We will demonstrate here that multimodality can face both missing and noisy modality issues in an elegant way. 

Multimodal models can fuse the multiple input data in various ways. Fusing data at the input-level, by projecting it onto a new dimensional space, is called Early fusion \cite{atrey_fusion_survey}. This has been the most common approach in previous sleep staging research. Such casting on a new space can be challenging especially if modalities differ significantly in dimensionality and sampling rate \cite{ramachandram2017deep}. Fusing on the decision-level by an ensemble of single-modality networks has been a more straightforward designing approach \cite{breiman1996bagging}, however, it reduces the multimodal collaboration. Such approaches are called Late fusion. Intermediate solutions include fusing the modalities before the classifier, referred to as Mid-Late fusion \cite{CHATZICHRISTOS2022341}. Despite being simplistic, this distinction based on the point of multimodal fusion seems to apply to the vast majority of multimodal architectures.

A further distinction of Late/Mid-Late models is whether they include the communication among the modality-specific parts of networks and the various ways that have been used for such communication. CoRe-Sleep, the multimodal network introduced in this paper, falls into this category allowing mid-late fusion to have coordinated representations \cite{baltruvsaitis2018multimodal}. Most works follow the general framework of aggregating information from one part of the network and transmitting it in a compressed form to the others. This can be achieved for example with the summation of intermediate CNN channels \cite{hazirbas2016fusenet}, the exchange of  intermediate representations \cite{wang2020channel_exchange}, any of the Squeeze and Excite gates \cite{Nagrani2021_AttBottleneck} or cross-attention weights \cite{lu2016hierarchical, tsai2019multimodal, li2022blip, li2021align, yu2022coca}. The latter approach will also be exploited by CoRe-Sleep. 

\subsection*{Our Contribution}

We propose CoRe-Sleep, a multi-modal model for sleep staging. The model is based on a TF backbone encoder \cite{vaswani2017attention, devlin2018bert}, the coordinated representations multimodal fusion and is trained with a multi-task objective. CoRe-Sleep's TF encoder uses an 8-layer of inner-outer block structure similar to SleepTransformer. The inner TF block processes the features extracted from Short-Time Fourier Transform (STFT) within a 30-second window, while the outer TF block maps interactions between the aggregated features of several sequential windows, 21 in this case (10.5 minutes) following \cite{phan2021xsleepnet, phan2022sleeptransformer}. Four TF blocks are used, one for each of the unimodal and multimodal processing of each modality. In the multimodal blocks, pairwise cross-attention is added as means of communication between modalities, similar to \cite{li2022blip, li2021align, yu2022coca}, to ensure the coordination of the modalities. The model is trained using multiple supervised losses for the unimodal and multimodal predictions, and a self-supervised objective for aligning the unimodal representations, following previous works in vision and text \cite{radford2022robust, li2022blip}.

The architectural choices and the training objective of CoRe-Sleep led to the four benefits emerging:

\begin{itemize}
\item State-of-the-art results when trained and tested with EEG and EOG modalities on the largest public dataset, the Sleep Heart Health Study (SHHS).
\item Robustness to missing modalities during 
inference; Improved performance compared to solely EEG-trained models and its own unimodal equivalent when tested with only one modality. Training with both modalities benefits unimodal testing.
\item Robustness to highly noisy data when it is present in one of the modalities.
\item Ability to train even with missing-modality data, i.e. allowing for the use of patient recordings that include only one modality.

\end{itemize}

The rest of the paper is organised as follows. In Section \ref{sec:method}, we present the dataset and the methods used in CoRe-Sleep and we introduce the experiments. In Section \ref{sec:results}, we describe and discuss the results, comparing our method with previous state-of-the-art and additional benchmark models. Lastly, we conclude the article in Section \ref{sec:conclusion}. 

\section{Data \& Methods}
\label{sec:method}

This section presents the architecture of the multimodal fusion network CoRe-Sleep and its multi-task objective. We begin by providing a brief overview of the largest publicly available dataset that is used in all our subsequent experiments.

\subsection{Data}
\label{sec:dataset}

 \textbf{SHHS} \cite{quan1997sleep, zhang2018national}. Initially intended to determine cardiovascular and other consequences of sleep-disordered breathing, SHHS consists of two rounds of recordings. We experiment with data from the first round, recorded between 1995-1998 (SHHS-1). It contains 5.791 subjects aged from 39 to 90 and includes several modalities.  We will use the EEG (C4-A1) sampled at 125Hz, and EOG (L-R) sampled at 50Hz. Scoring was completed using the Rechtschaffen and Kales (R\&K) guideline \cite{rechtschaffen1968manual}. Following previous work, we merge stages N3 and N4 into the N3 stage, while Movement and Unscored labels were discarded.

To preprocess the data a series of steps have been followed. Initially, we load the signals and labels. Following previous works \cite{phan2021xsleepnet}, we discard patients' recordings that do not contain windows with labels from each of the five classes. We also discard segments from the beginning and the end of the recording if the number of wake stage windows is greater than the windows of another stage. More specifically we discard  $\frac{\#wake-\#2^{nd} dominant}{2}$ from each side. EEG and EOG are resampled at 100Hz and filtered with a bandpass FIR on [0.3, 40] Hz and on [0.3, 23] Hz. The signals are transformed with STFT using 2-second time bins with 1-second overlap and 256 points hamming window, resulting in 128-dimension features. Then we create the 30-second windows without overlap with their corresponding label, using the majority policy when more than one class is presented within the 30-seconds. Finally, we split the patients of the dataset using a random split of 70-30\% for the train-test set and from the training we keep 100 patient recordings for the validation set.

\begin{figure*}[!t]
    \begin{subfigure}[t]{.49\textwidth}
        \centering   
        \includegraphics[width=0.8\textwidth]{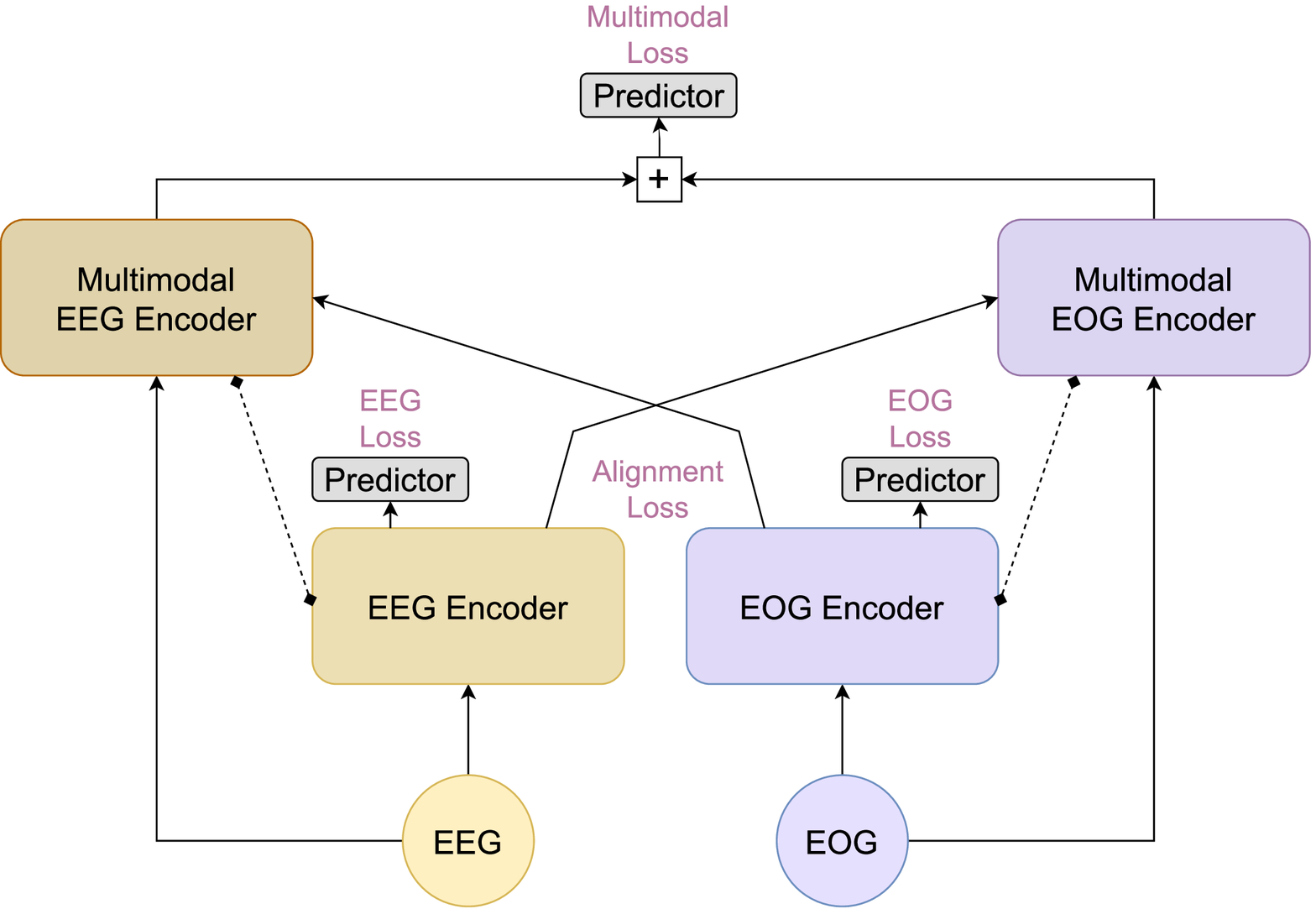}
        \caption{CoRe-Sleep General Layout}
        \label{fig:general_layout}
    \end{subfigure}
    \begin{subfigure}[t]{.49\textwidth}
        \centering
        \includegraphics[width=0.8\textwidth]{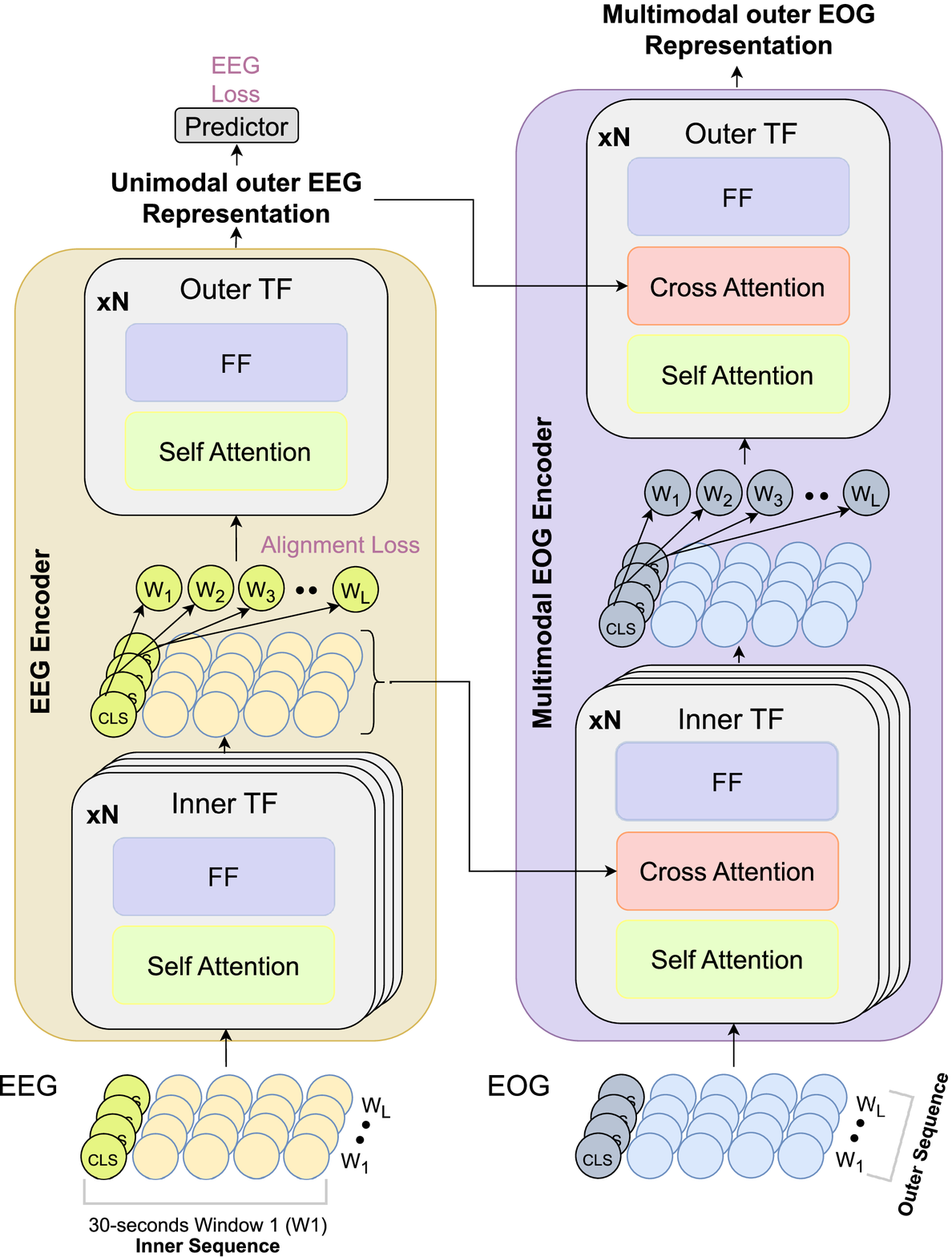}
        \caption{Unimodal and Multimodal Encoders}\label{fig:uni_mult_encoder}
    \end{subfigure}
    
\caption{CoRe-Sleep: Coordinated Representation -based multimodal fusion model for Sleep Staging \textbf{a)}: The general layout of the model is illustrated. Each modality (Electroencephalography (EEG) and Electroculography (EOG)) is processed by its unimodal encoder to create the corresponding unimodal representations, which are then fed to the multimodal encoder. The multimodal encoders cross-attend with the unimodal representation of the other modality, creating the multimodal representation. Supervised losses are calculated based on each representation (EEG, EOG, and multimodal). The model also uses a self-supervised alignment loss between the unimodal representations, as in \cite{radford2021learning}. The dashed lines represent the shared weight connections of the Feed-Forward (FF) and the Self-Attention weights. \textbf{b)}: The figure shows a unimodal and a multimodal encoder with their components. The input data has both inner and outer dimensions, which are processed by their respective N-layers Transformer (TF) blocks. To summarize the inner sequence, a learnable class embedding {\small [CLS]} is utilized. The outputs of the inner and outer TF of one modality are used to cross-attend the multimodal representation of the other modality, leading to the coordinated representations.}
\label{fig:coresleep_model}
\end{figure*}

\subsection{Transformer Backbone}
\label{sec:tf_backbone}
The TF backbone is the TF encoder described in \cite{vaswani2017attention}. The encoder is responsible for processing the input sequence. The TF encoder consists of multiple layers, each with two components. The first is the multi-head self-attention (SA), which allows different parts of the input sequence to interact. The second is a fully connected feed-forward (FF) network, which applies a non-linear transformation to each position in the sequence. Both components have a residual connection and a (post-) layer normalization \cite{ba2016layer}. A distinctive TF encoder is utilized to handle each modality $X_m$. The equations describing this encoder are:

\scalebox{0.95}{\parbox{\linewidth}{%
 \begin{align}
        att_h(X_m) &= \text{softmax}(\frac{W^K_h X_m (W^Q_h X_m)^\intercal}{\sqrt{d_k}})W^V_h X_m\\
        SA(X_m) &= \text{concat}[att_1(X_m), .., att_H(X_m)]W^O \\
        Z_m &= \text{layernorm}(X_m+SA(X_m)) \\
        FF(Z_m) &= max(0, Z_m W^{F}_1 + b^{F}_1) W^{F}_2 + b^{F}_2 \\
        output &= \text{layernorm}(Z_m+FF(Z_m))      
\end{align}
}}

Input modality $X_m \in \mathbb{R}^{batch \times sequence \times features}$ is a batch of sequence of features. The weights of the SA component have dimensions $W^K_h, W^Q_h \in \mathbb{R}^{d_{model} \times d_k}$, $W^V_h \in \mathbb{R}^{d_{model} \times d_u}$, $W^O \in \mathbb{R}^{Hd_u \times d_{model}}$ and $d_{model}=128$, $H=8$, $d_{k}=16$, $d_{u}=128$. The SA allows for measuring pairwise similarity between all the feature vectors in the sequence, re-weighting them to favour those with greater similarity. The utilization of multi-head facilitates the extraction of numerous interactions among different parts of the sequence. For the FF a two-layer multi-layer perception (MLP) with the Rectified Linear Unit (ReLU) \cite{nair2010rectified} as activation function has been used, with weights $W^{F}_1 \in \mathbb{R}^{d_{model}}$, $W^{F}_2 \in \mathbb{R}^{ d_{ff} \times d_{model}}$, biases $b^{F}_1 \in \mathbb{R}^{d_{ff}}$, $b^{F}_2 \in \mathbb{R}^{d_{model}}$ and dimensionality $d_{ff}=1024$. The point-wise FF is responsible to process each feature vector of the sequence independently of the others.

\subsection{CoRe-Sleep Architecture}

CoRe-Sleep's architecture utilizes the TF backbone and the multimodal fusion of coordinated representations. Initially, we describe the encoder for a single modality and subsequently, we proceed with the multimodal part of the network.

In CoRe-Sleep architecture, the inner-outer framework described in Section \ref{sec:introduction} is used, similar to \cite{phan2021xsleepnet, tinysleepnet, phan2022sleeptransformer}. Each modality has its own unimodal encoder which is an 8-layer TF encoder. The first 4-layers are processing the inner sequence, finding the interactions between the features extracted from STFT within the 30-second windows. Afterwards, the inner sequence features are aggregated. The last 4-layers of the TF encoder process the outer sequence, the interactions between these aggregated features of neighbouring windows. We allow for maximum 21 sequential windows, following \cite{phan2021xsleepnet, phan2022sleeptransformer}, but recent works indicate that even greater numbers could be beneficial \cite{L-SeqSleepNet}. The Inner-Outer scheme is visualised in Figure \ref{fig:uni_mult_encoder}.  

CoRe-Sleep's unimodal encoder is similar to SleepTransformer \cite{phan2022sleeptransformer} with two modifications: 
\begin{enumerate}
  \item It uses a learnable embedding, for the class {\small [CLS]}, for aggregating inner-sequence features instead of a separate attention module. This allows more interaction via multi-head attention and multi-layer aggregation.
  \item It uses learnable relative positional embeddings \cite{shaw2018self} instead of absolute sinusoidal ones. These allow the model to learn dependencies between positions in the feature sequence and are more efficient because they are incorporated into the attention mechanism.
\end{enumerate}
These changes result in a small improvement. The unimodal encoder's output is utilized by a predictor and together they form the unimodal equivalent network, which is discussed in later sections.

To achieve multimodal fusion, we exploit coordinated representations. Coordinated representation is a mid-late fusion scheme, where each modality has a separate network stream, with established communication between them. We first calculate the inner and outer sequence feature representations of each modality with the modality-specific unimodal encoders, described above. We feed then each modality's inner and outer representation to the corresponding multimodal encoder branches of the other modality, Figure \ref{fig:general_layout}. The weights of SA and FF are shared among the multimodal and the unimodal encoders of each modality. The multimodal encoders have the same structure as the unimodal encoders with an additional cross-attention (CA) component to let one modality affect the representation of the other, similar to \cite{vaswani2017attention, li2022blip, li2021align, yu2022coca, tsai2019multimodal}. The CA component allows the pairwise similarities between the sequence steps of the two modalities $X_1, X_2$ to affect the importance by re-weighting one of them. The equations describing the CA component are:
\begin{align}
    \scalebox{0.95}{
    $
    \text{c-att}_h(X_1| X_2) = \text{softmax}(\frac{W^K_h X_1 (W^Q_h X_2)^\intercal}{\sqrt{d_k}})W^V_h X_1 
    $
    } 
    \\
    \scalebox{0.93}{$
        \text{CA}(X_1|X_2) = \text{concat}[\text{c-att}_1(X_1|X_2), .., \text{c-att}_H(X_1|X_2)]W^O 
    $}
\end{align}

 The matrices $W^K_h$, $W^Q_h$, $W^V_h$, $W^O$ have the same dimensionality as in the TF encoder described in \ref{sec:tf_backbone}, without being the same set of parameters. The CA component is placed between the SA and the FF block of each layer. Multiple sequential CAs can be used to extend CoRe-Sleep on more than two input modalities. The output of each multimodal encoder, which is dedicated to one modality, can be considered the representation of the corresponding modality grounded on (influenced by) the remaining modalities. The final representation based on all modalities is a summation of the different grounded multimodal representations. Such summation allows all predictors to have the same number of parameters and possibly share them among the predictors of each modality. 

\subsection{Training Objectives}

The model jointly optimizes three distinct loss functions, see Figure \ref{fig:general_layout}. The two of them are supervised Cross-Entropy (CE) losses, from the predictions of the unimodal and the multimodal network's output. The third is a self-supervised alignment (AL) loss between the two modalities. 

\textbf{Alignment Loss (AL)}: It exploits the fact that EEG and EOG signals are measured at the same time, allowing to cast this as a self-supervised task of predicting which 30-second windows of one modality corresponds to the other modality. This is motivated by literature on vision-text integration \cite{radford2021learning, li2022blip, li2021align, yu2022coca}, where images align with captions and the task is to match them. CoRe-Sleep compares each window's unimodal representation with those of other modalities to predict matching pairs, with each representation being dedicated to a single window. We cast the problem as a double classification task with each modality having to predict its position in the other one, similar to \cite{radford2021learning}. 

\textbf{Multi-Supervised Loss (MS)}: CoRe-Sleep's architecture creates both unimodal and multimodal representations. MS loss is the summation of supervised losses arising from placing additional predictors to each unimodal representation. It reduces the competing effects of each modality \cite{huang2022modality} by allowing the worse-performing modality to affect proportionally the total loss. MS loss enforces the network to be able to perform well with all modalities present as well as with each modality separately. In CoRe-Sleep the multiple predictions come at minimal computational overhead.

Given a training set of patients full night recordings, the goal is to identify the sleep label $y$ for each 30-second window. The patient recordings are segmented in windows to create the dataset $X = (\{X^1_{1}, .. X^1_{M}\}, .., \{X^L_{1}, .. X^L_{M}\})$ where $M$ denotes the number of modalities, in this paper $M=2$ and $L$ the total number of windows in the dataset. Each $X^l_{m} \in \mathbb{R}^{T \times D}$ and $m=1, .., M$ contains the frequency features with resolution $D=128$ extracted from STFT where a smaller window of 2 seconds with a 1-second overlap is applied. It results in $T=29$ feature vectors representing a 30-second window. 

Sequential windows include useful information therefore we preserve such information by processing multiple neighbouring windows together. Following the inner-outer framework, we process first the within-window features $\{X^l_1, .., X^l_M\} \in \mathbb{R}^{M\times T \times D}$, aggregating information on the output of the inner $\in \mathbb{R}^{M\times 1 \times D}$. Subsequently, we map the interaction between neighbouring windows on the outer sequence. To achieve that we take sequential windows of the same recording, allowing for a maximum of 21 sequential windows following \cite{phan2021xsleepnet, phan2022sleeptransformer}. Each model $f$ trained in the experiments dedicates a subpart of its total parameters $\theta$ for each of the modalities $m$, denoted as $\theta_{m}$. Each such subpart of the network calculates the unimodal predictions used to calculate the MS loss. In conclusion following \cite{radford2021learning}, the alignment loss is obtained by summing up the CE losses, where the predictions are calculated using the outer product of pairwise unimodal representations, and the targets are set to the identity matrix $I$. Intuitively AL tries to predict which window of one modality belongs to the other. The target of AL loss is the outer sequence length, so the AL loss has a different magnitude than the rest of the losses, therefore we include a scaling parameter $\lambda_{A}=0.1$. We observed model performance is not sensitive to small changes on $\lambda_{A}$; however this still might not be the optimal value.    

The final training objective $L( X, y / f_{\theta})$ is a sum of the previously mentioned:
\begin{align}    
        L( X, y | f_{\theta}) &= \text{CE}( f_{\theta}( X), y)\text{+}\text{MS}(X, y | f_{\theta}) + \text{AL}(X | f_{\theta}) \\
        \text{with}\quad&MS(X, y | f_{\theta})) = \sum_{m=1}^M CE( f_{\theta_{m}}( x_i ), y) \quad\text{and}\\
        AL(X | f_{\theta})) &= \lambda_{A} \sum_{m=1}^M \sum\limits_{\substack{j=1 \\ m\neq j}}^M \text{CE}(f_{\theta_{m}}(X_m)f_{\theta_{j}}(X_j)^\intercal, I) 
\end{align}

\begin{table*}
\renewcommand{\arraystretch}{1.3}
\caption{
The table compares the unimodal and multimodal performance of the Core, Early, and Mid-Late architectures on the SHHS-1 dataset. Systematically AL and MS losses are added until Core-Sleep achieves the best results overall. All models are trained on EEG \& EOG and tested on EEG (middle), EOG (right) or both (left). We perform training three times with different dataset splits to report variance. Results from the literature on this dataset are added below the dashed line. Metrics used: Accuracy, Cohen's Kappa, Macro-F1.}
\label{table:shhs}
\centering
\begin{tabular}{llll|lll|lll}\toprule
Dataset: \textbf{SHHS-1} &  \multicolumn{9}{c}{Overall Metrics}
 \\\cmidrule(lr){2-10}
Test Modalities & \multicolumn{3}{c}{EEG-EOG} & \multicolumn{3}{c}{EEG} & \multicolumn{3}{c}{EOG} \\
\cmidrule(lr){2-4}\cmidrule(lr){5-7}\cmidrule(lr){8-10}
Model & Acc  & $\kappa$ & MF1& Acc  & $\kappa$ & MF1& Acc  & $\kappa$ & MF1\\
\toprule
Early 
& 89.1 {\tiny$\pm$0.0} & 0.847 {\tiny$\pm$0.001} & \textbf{81.7 {\tiny$\pm$0.3}}
& 58.0 {\tiny$\pm$1.4} & 0.403 {\tiny$\pm$0.025} & 42.5 {\tiny$\pm$1.9}
& 43.6 {\tiny$\pm$10.3} & 0.201 {\tiny$\pm$0.111} & 29.2 {\tiny$\pm$8.1}\\
Mid-Late 
& 89.1 {\tiny$\pm$0.1} & \textbf{0.848 {\tiny$\pm$0.002}} & 81.6 {\tiny$\pm$0.2}
& \textbf{85.4 {\tiny$\pm$0.4}} & \textbf{0.797 {\tiny$\pm$0.006}} & \textbf{78.2 {\tiny$\pm$0.7}}
& \textbf{75.4 {\tiny$\pm$2.3}} & \textbf{0.639 {\tiny$\pm$0.039}} & \textbf{59.1 {\tiny$\pm$2.5}} \\
CoRe 
& 89.1 {\tiny$\pm$0.0} & 0.847 {\tiny$\pm$0.001} & 81.6 {\tiny$\pm$0.3} 
& 64.7 {\tiny$\pm$8.4} & 0.458 {\tiny$\pm$0.139} & 46.2 {\tiny$\pm$13.1} 
& 28.3 {\tiny$\pm$13.2} & -0.016 {\tiny$\pm$0.126} & 11.9 {\tiny$\pm$6.2}  \\
 \cline{1-10} 

Early \textbf{+ AL} 
& 89.2 {\tiny$\pm$0.0} & 0.849 {\tiny$\pm$0.000} & \textbf{81.9 {\tiny$\pm$0.2}}
& 49.7 {\tiny$\pm$14.4} & 0.298 {\tiny$\pm$0.179} & 39.4 {\tiny$\pm$13.8}
& 34.0 {\tiny$\pm$4.7} & 0.094 {\tiny$\pm$0.066} & 21.6 {\tiny$\pm$9.1}\\
Mid-Late  \textbf{+ AL} 
& 89.2 {\tiny$\pm$0.1} & 0.848 {\tiny$\pm$0.002} & 81.7 {\tiny$\pm$0.3} 
& \textbf{85.7 {\tiny$\pm$0.3}} & \textbf{0.800 {\tiny$\pm$0.004}} & \textbf{78.2 {\tiny$\pm$0.6}} 
& \textbf{74.8 {\tiny$\pm$2.3}} & \textbf{0.627 {\tiny$\pm$0.036}} & \textbf{57.2 {\tiny$\pm$3.0}}\\
CoRe \textbf{+ AL} 
& \textbf{89.3 {\tiny$\pm$0.2}} & \textbf{0.850 {\tiny$\pm$0.002}} & 81.8 {\tiny$\pm$0.1}
& 53.7 {\tiny$\pm$7.5} & 0.341 {\tiny$\pm$0.120} & 39.6 {\tiny$\pm$11.0} 
& 43.9 {\tiny$\pm$4.1} & 0.172 {\tiny$\pm$0.069} & 23.0 {\tiny$\pm$3.1}\\ 
 \cline{1-10}   
 
Early \textbf{+ MS} 
& \textbf{89.4 {\tiny$\pm$0.1}} & \textbf{0.851 {\tiny$\pm$0.002}} & \textbf{82.1 {\tiny$\pm$0.3}}
& 81.7 {\tiny$\pm$1.1} & 0.742 {\tiny$\pm$0.015} & 65.8 {\tiny$\pm$1.0}
& 77.7 {\tiny$\pm$1.5} & 0.686 {\tiny$\pm$0.018} & 62.4 {\tiny$\pm$1.2}\\ 
Mid-Late  \textbf{+ MS} 
& 89.2 {\tiny$\pm$0.1} & 0.849 {\tiny$\pm$0.002} & 81.6 {\tiny$\pm$0.1} 
& 87.7 {\tiny$\pm$0.2} & 0.828 {\tiny$\pm$0.003} & 80.1 {\tiny$\pm$0.2} 
& \textbf{84.9 {\tiny$\pm$0.2}} & \textbf{0.787 {\tiny$\pm$0.002}} & 74.4 {\tiny$\pm$0.2}\\
CoRe \textbf{+ MS} 
& \textbf{89.4 {\tiny$\pm$0.1}} & \textbf{0.851 {\tiny$\pm$0.002}} & 82.0 {\tiny$\pm$0.3} 
& \textbf{87.9 {\tiny$\pm$0.2}} & \textbf{0.830 {\tiny$\pm$0.003}} & \textbf{80.3 {\tiny$\pm$0.4}} 
& \textbf{84.9 {\tiny$\pm$0.1}} &\textbf{0.787 {\tiny$\pm$0.002}} & \textbf{74.5 {\tiny$\pm$0.2}}\\
 \cline{1-10} 
 
Early \textbf{+ MS + AL} 
& \textbf{89.5 {\tiny$\pm$0.1}} & \textbf{0.853 {\tiny$\pm$0.002}} & \textbf{82.3 {\tiny$\pm$0.2}}
& 87.1 {\tiny$\pm$1.7} & 0.820 {\tiny$\pm$0.021} & 79.7 {\tiny$\pm$1.6}
& 83.4 {\tiny$\pm$3.0} & 0.770 {\tiny$\pm$0.036} & 74.0 {\tiny$\pm$2.3} \\
Mid-Late \textbf{+ MS + AL} 
& 89.3 {\tiny$\pm$0.1} & 0.851 {\tiny$\pm$0.002} & 81.9 {\tiny$\pm$0.1} 
& \textbf{88.0 {\tiny$\pm$0.2}} & 0.831 {\tiny$\pm$0.003} & 80.4 {\tiny$\pm$0.2} 
& \textbf{85.2 {\tiny$\pm$0.1}} & \textbf{0.792 {\tiny$\pm$0.002}} & 75.1 {\tiny$\pm$0.1}  \\
CoRe-Sleep (\textbf{MS + AL}) 
& \textcolor{blue}{\textbf{89.5 {\tiny$\pm$0.1}}} & \textcolor{blue}{\textbf{0.853 {\tiny$\pm$0.002}}} & \textcolor{blue}{\textbf{82.3 {\tiny$\pm$0.3}} }
& \textcolor{blue}{\textbf{88.2 {\tiny$\pm$0.2}}} & \textcolor{blue}{\textbf{0.834 {\tiny$\pm$0.003}}} & \textcolor{blue}{\textbf{80.8 {\tiny$\pm$0.4}} }
& \textcolor{blue}{85.3 {\tiny$\pm$0.1}} & \textcolor{blue}{0.792 {\tiny$\pm$0.001}} & \textcolor{blue}{75.3 {\tiny$\pm$0.3}}\\

Unimodal Equivalent & - & - & -  & 87.7 {\tiny$\pm$0.2} & 0.828 {\tiny$\pm$0.003} & 80.1 {\tiny$\pm$0.4} & 85.1 {\tiny$\pm$0.1} & 0.790 {\tiny$\pm$0.001} & 75.0 {\tiny$\pm$0.2}\\
\hdashline
XSleepNet \cite{phan2021xsleepnet} & 88.8 & 0.843 & 82.0  & 87.6 & 0.826 & 80.7  & - & - & -\\
SleePyCo \cite{lee2022sleepyco} & - & - & -  & 87.9 & 0.830 & 80.7 & - & - & -\\
SleepTransformer \cite{phan2022sleeptransformer} & - & - & - & 87.7 & 0.828 & 80.1 & - & - & -\\
\bottomrule
\end{tabular} 
\end{table*}

\subsection{Architecture Benchmarks (Early \& Mid-Late Fusion)}

\begin{figure}[!t]
    \begin{subfigure}[t]{.49\linewidth}
        \centering   
        \includegraphics[width=0.9\textwidth]{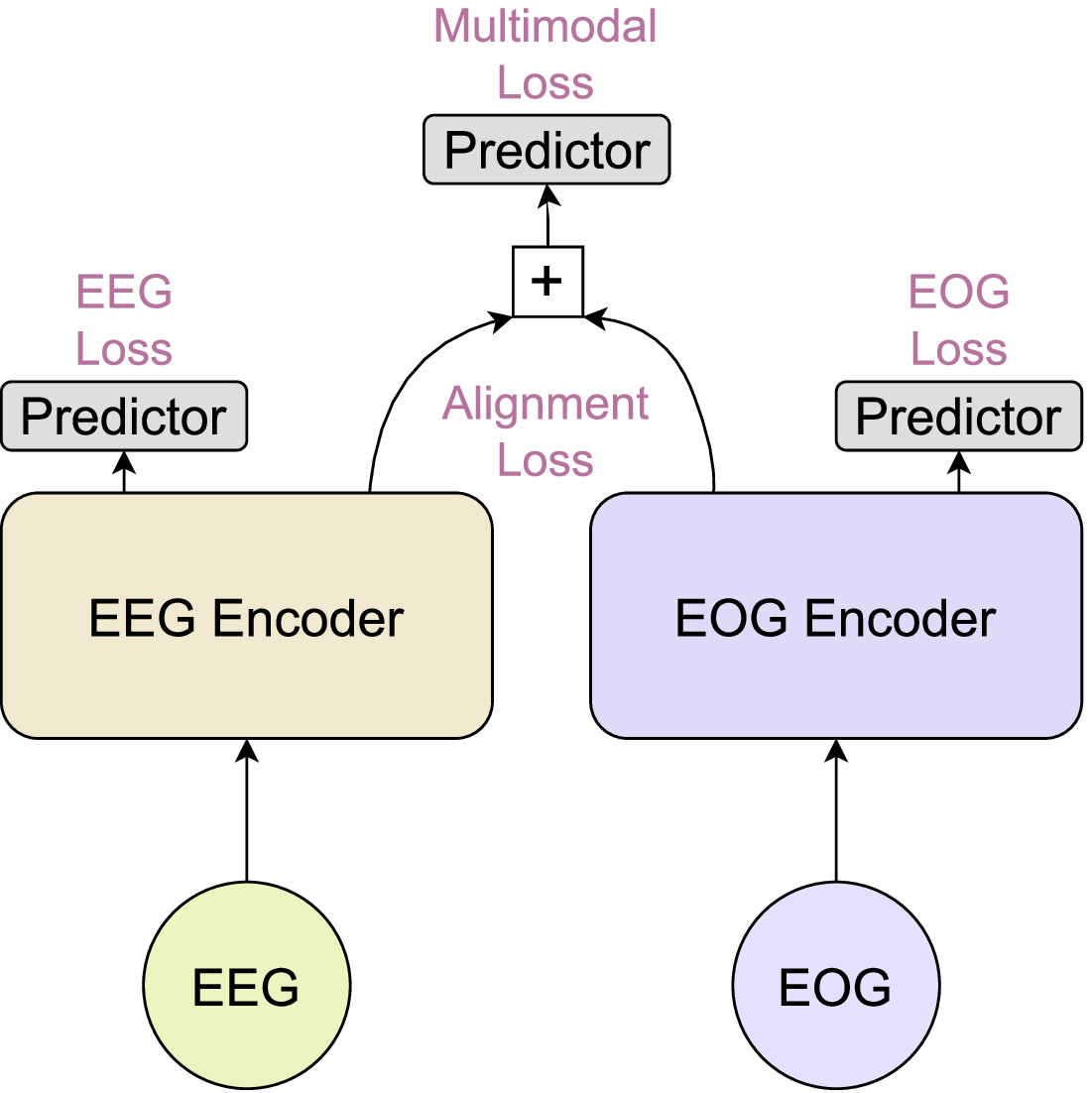}
        \caption{Mid-Late Fusion}
        \label{fig:late_fusion}
    \end{subfigure}
    \begin{subfigure}[t]{.49\linewidth}
        \centering   
        \includegraphics[width=0.71\textwidth]{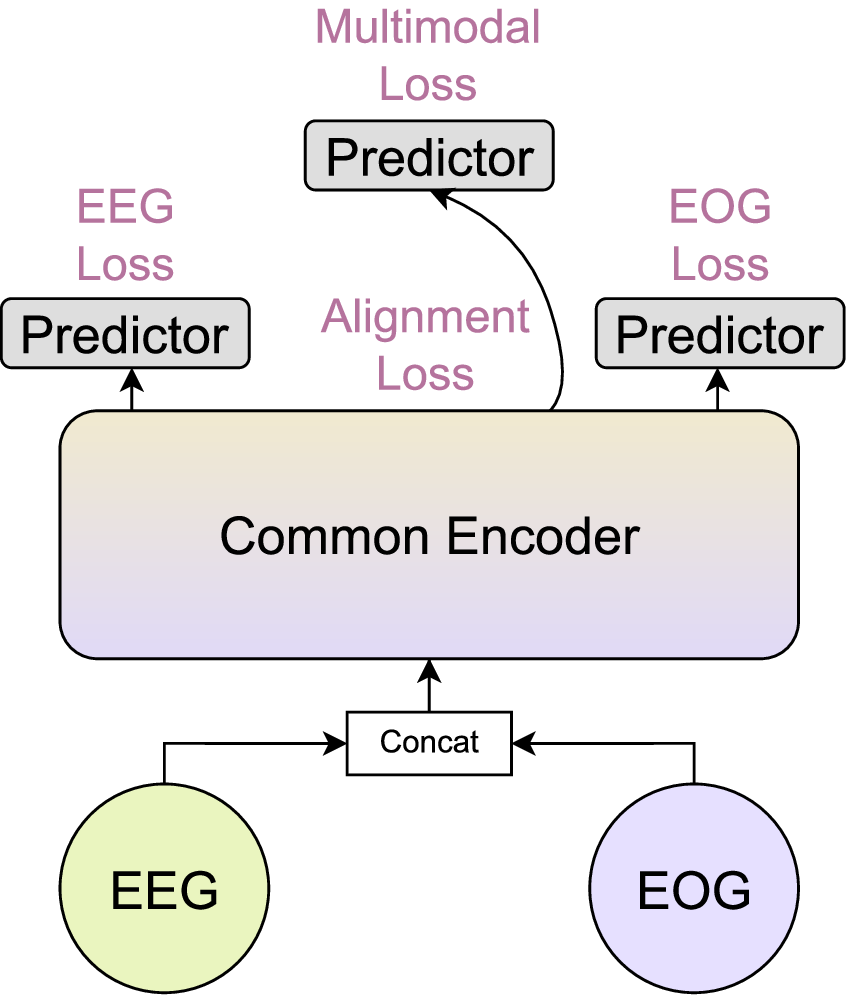}
        \caption{Early Fusion}
        \label{fig:early_fusion}
    \end{subfigure}
\caption{
\textbf{a)} Mid-Late fusion scheme (MS + AL): Each modality has its own encoder and information gets only merged before the multimodal predictor. The Mid-Late model refers to the model trained solely on the Multimodal loss, while the models incorporating MS and AL are explicitly identified. Similar to CoRe-Sleep, this Mid-Late fusion scheme can derive both unimodal and multimodal predictions in a single forward pass.
\textbf{b)} Early fusion scheme (MS + AL): In the Early model, both EEG and EOG embeddings are processed by a shared encoder to calculate the multimodal loss. The Early model is defined again without the MS and AL losses. Unlike the Mid-Late scheme, the Early model requires two forward passes to compute the unimodal representations.}
\label{fig:early_late_schema}
\end{figure}

We introduce here two benchmark models which along with previous state-of-the-art from literature we compare with CoRe-Sleep. These models assist in illustrating the benefits of the multimodal fusion architecture and the training objectives. We take two of the uttermost cases, the Early and the Mid-Late fusion. As the name suggests, the former fuses multimodal input before encoding and the latter trains modality encoders separately and only fuses them before the predictor, see Figure \ref{fig:early_late_schema}. 

\textbf{Early fusion} is a model structure that has no modality-specific processing and data are merged (by concatenation) before the model in the data-level. Only preprocessing steps are distinctive between the two modalities. The same TF blocks are processing both modalities by considering them parts of the sequence both in inner and outer parts. That allows each modality to attend to each other and creates the fonts to identify better multimodal interactions. TF distinguish the time relationships with the positional encoding. Similarly, to achieve the discrimination between the two modalities we sum another learnable embedding unique for every modality \cite{ma2022multimodal}. We note that Early fusion requires more than one forward pass to obtain the unimodal representations, hence reducing the inference speed.

\textbf{Mid-Late fusion} is the architecture where each modality has its own unimodal encoder which produces the unimodal representations and by just summing them up we get the multimodal as well. Calculating both the unimodal and alignment losses comes straightforward since modalities are disentangled. We define Mid-Late fusion as the model architecture without these additional losses. Finally, both unimodal and multimodal predictions can be acquired with a single forward pass.

On top of these benchmark models, we define CoRe, the model with the same architecture as CoRe-Sleep but trained without the additional training objectives MS and AL. CoRe follows the Mid-Late model of creating a separate encoder for each modality but with established communication between them. CoRe, Early and Mid-Late models have the same TF backbone and inner-outer logic in each of their blocks. In all models, the final multimodal representation is created by summing the two unimodal representations. A learnable class embedding {\small [CLS]} is utilised to aggregate inner-sequence information, following \cite{devlin2018bert}. Each model is trained on the multimodal predictor CE loss and the two additional losses (MS and AL) are gradually added to show their benefit.

\subsection{Implementation}

All our models are implemented in PyTorch \cite{paszke2017automatic} and they have been trained on a single GPU node. We use Adam \cite{kingma2014adam} optimizer with the learning rate and the weight decay being 1e-4. We train with batch size 16 and outer sequence 21 \cite{phan2022sleeptransformer}, resulting in 336 sleep labels per batch. We validate our models every 400 optimization steps. We also use a cosine annealing \cite{loshchilov2016sgdr} scheduler with a maximum learning rate of 0.03 and 20k warm-up optimization steps. We consider a model converged when it has not been improved in the last 100k optimization steps (approximately 9 epochs) and then we apply early stopping. 

Each unimodal and multimodal encoder has an inner and outer block of TF, each with 4 layers of post-normalization TF. Relative positional embedding \cite{shaw2018self} is only added to key vectors in attention components. Models have 128-dimensional input, 8 attention heads, 1024-dimensional MLP for FF, and 2-layer MLP for predictors all with 0.3 dropout probability. Weights are shared among the SA and FF parts of the unimodal and multimodal encoders.

\subsection{Experiments}

A series of experiments demonstrate the improvements of CoRe-Sleep. The importance of each component is evaluated by comparing the performance of Early, Late, and CoRe when trained and tested on multimodal (EEG and EOG) input, and gradually AL and MS losses are added. We note that CoRe-Sleep is the CoRe architecture with added MS and AL losses. Next, we test the models' performance when one of the modalities (EEG or EOG) is missing with the goal to assess the robustness of missing modalities. For these experiments, the models do not need to be retrained. Only the unimodal equivalent model and the models reported in previous studies are trained and tested using one modality. 

To explore the robustness of the model to noisy modalities, a subset of SHHS-1, where one of the modalities is highly noisy, is extracted. Such noisy cases can be identified by thresholding the difference in STD of EEG and EOG modalities. The STD is evaluated on large chunks of 10 minutes to avoid picking artefacts. Only the patients that present corruption in over 40$\%$ of the recording are kept, to avoid picking patients with noise segments at the beginning or end of a recording where sensors might not yet be attached or have been detached. This subsampling is applied three times, once for each split of the dataset with [17, 20, 11] patients having over 40\% noisy EEG and [10, 4, 3] having over 40\% noisy EOG on the different spits. Notably, EEG, the dominant modality for sleep staging, appears to contain more long noisy periods. We test the models on these sets of patients to evaluate robustness to noisy modalities.

CoRe-Sleep is capable of also being trained with modality-incomplete data. The motivation of this last experiment is to examine the behaviour of the model under such unimodal data additions. CoRe-Sleep is trained from scratch with an initial subset of 100 multimodal patients and a variable number of unimodal ones (either from EEG, EOG or both). Only on the initial 100, are the MS and AL losses calculated. When both unimodal modalities are added they don't belong to the same patient. The experiment is repeated three times adding gradually patients with EEG, EOG or from both modalities.

\begin{figure*}[t]
\centering
\includegraphics[width=\textwidth]{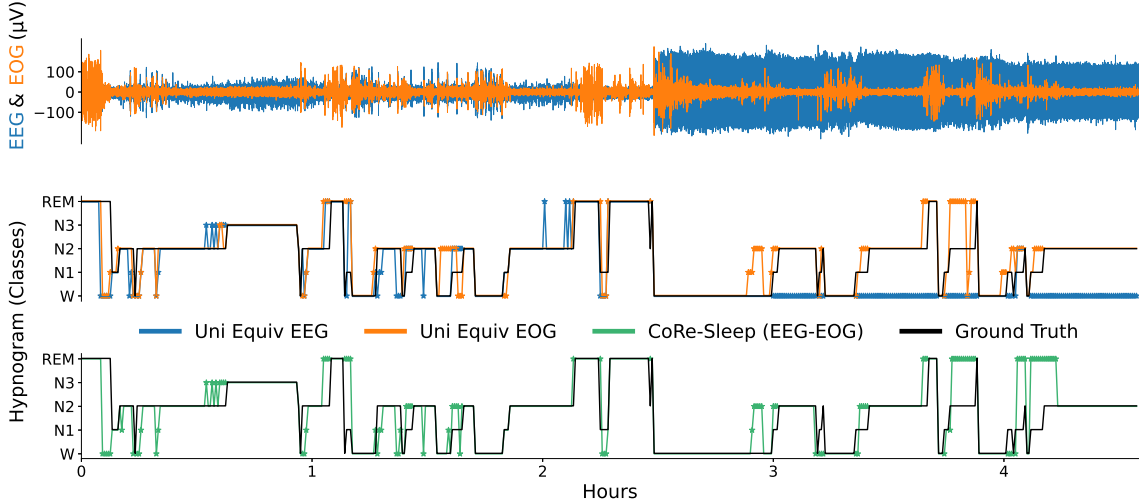}
\caption{This figure provides an illustration of a patient from SHHS-1 whose EEG modality was highly corrupted in the second half of the recording. The top section of the figure displays the EEG (C4-A1) and EOG (L-R) signals, with a sudden increase in the standard deviation of the EEG signal indicating the presence of noise. The two rows below show the class predictions (hypnogram) of the unimodal EEG and EOG models, as well as the multimodal EEG-EOG CoRe-Sleep model. Notably, in the highly noisy area, the EEG model performed poorly, while the multimodal model remained robust by leveraging the information provided by the EOG.}
\label{fig:time_std_preds_patient}
\end{figure*}
\section{Results \& Discussion}
\label{sec:results}
This section demonstrates the contribution of each component in the multimodal fusion in standard, missing-modality and noisy scenarios, and analyses the behaviour of our model trained on incomplete multimodal data.

\subsection{Multimodal}
Results on training and testing CoRe-Sleep and the benchmark models (Early and Mid-Late) with multimodal input show that optimizing the multimodal fusion leads to outperforming previous state-of-the-art by a small margin, as shown in the first column of Table \ref{table:shhs}. All models, namely Early, Mid-Late and Core, performed similarly when no additional losses were added. However, adding the AL loss improved the performance of Early and Core, which are the models that include attention interactions between the two modalities. MS loss contributes consistently to all three models. We demonstrate that CoRe-Sleep performs similarly to Early ( when MS + AL are added), which is the model that allows the maximum interactions between modalities. 

\subsection{Missing Modality}
\label{sec:missing_modality}

One of the main advantages of CoRe-Sleep is its ability to handle missing modalities. In this second experiment, the already trained models are evaluated in scenarios where only one of the two modalities is present. The second and third columns of Table \ref{table:shhs} display the missing modality scenario experimental results. Mid-Late fusion performs the best without the additional losses. This is expected, the model which has fewer interactions between modalities and each of its encoders is focused on a single modality, to not be affected as much absence of one modality. 

We observe how the models behave when extra losses are included. Each model shows a distinct response to the addition of AL loss. In Mid-Late it has a small contribution. In CoRe, when used without MS, it assists in balancing the performance between the two modalities. AL in the Early model deteriorates both the performances of EEG and EOG. However, when combined with MS, all models exhibit improved performance. Meanwhile, MS loss has a consistent effect on all models, indicating that each modality has to perform on each own as well. All models benefit from MS loss, with more effect on CoRe and Early. CoRe-Sleep's performance exceeds both in scenarios of missing EEG or EOG over all the other models. It even slightly outperforms its unimodal equivalent and the previous state-of-the-art models that have been trained solely on the present modality. 

The contribution of this experiment is twofold. Firstly, it shows that the multimodal fusion and the multi-task objective can strengthen the model's ability to perform inference with a subset of the modalities given during training. This addition makes CoRe-Sleep able to overcome issues such as sensor detachment, device deficiency or lack of all recording modalities. Secondly, the results show that, as CoRe-Sleep multimodal training surpasses its unimodal equivalent models both for EEG and EOG, unimodal models in general can benefit from multimodal training.

\begin{table}[t]
\renewcommand{\arraystretch}{1.3}
\caption{Comparison of model performance (Accuracy, Cohen's $\kappa$ and Macro-F1) on subset of Noisy SHHS-1 Patients}
\label{table:noisy_shhs}
\centering
\begin{tabular}{l|lll}\toprule
Dataset: Noisy SHHS-1 & \multicolumn{3}{c}{Overall Metrics}
 \\\cmidrule(lr){2-4}
Test Modalities & \multicolumn{3}{c}{EEG-EOG Noisy} \\
\cmidrule(lr){2-4}
Model & Acc  & $\kappa$ & MF1\\
\toprule

Unimodal Equivalent EEG
& 56.1 {\tiny$\pm$0.019} & 0.351 {\tiny$\pm$0.029} & 43.7 {\tiny$\pm$3.5}\\
Unimodal Equivalent EOG
& 80.6 {\tiny$\pm$2.1} & 0.722 {\tiny$\pm$0.03} & 69.5 {\tiny$\pm$2.0}\\ 

\midrule

Early
& 77.9 {\tiny$\pm$2.9} & 0.669 {\tiny$\pm$0.046} & 68.6 {\tiny$\pm$4.4}\\
Mid-Late
& 81.7 {\tiny$\pm$0.3} & 0.702 {\tiny$\pm$0.007} & 73.9 {\tiny$\pm$0.4}\\
CoRe
& 79.5 {\tiny$\pm$1.4} & 0.706 {\tiny$\pm$2.1} & 67.7 {\tiny$\pm$1.8} \\ 

\midrule

Early \textbf{+ AL}
& 81.1{\tiny$\pm$1.2} & 0.730{\tiny$\pm$0.017} & 70.4{\tiny$\pm$1.1}\\ 
Mid-Late \textbf{+ AL}
& 82.2{\tiny$\pm$0.7} & 0.746{\tiny$\pm$0.009}& 70.2{\tiny$\pm$0.7} \\ 
CoRe \textbf{AL}
& 82.2{\tiny$\pm$0.3} & 0.745{\tiny$\pm$0.006}& 71.4{\tiny$\pm$0.5} \\ 

\midrule

Early \textbf{+ MS} 
& 81.1{\tiny$\pm$1.0}  & 0.730{\tiny$\pm$0.015} & 70.2{\tiny$\pm$1.2}\\ 
Mid-Late \textbf{+ MS}
& 84.2{\tiny$\pm$0.5}  & 0.774{\tiny$\pm$0.008} & 73.2{\tiny$\pm$1.0}\\ 
CoRe \textbf{+ MS}
& 82.9{\tiny$\pm$1.1}  & 0.754{\tiny$\pm$0.015} & 71.5{\tiny$\pm$0.9}\\

\midrule

Early \textbf{+ MS + AL} 
& 83.1 {\tiny$\pm$0.5} & 0.758 {\tiny$\pm$0.007} & 72.2 {\tiny$\pm$0.6}\\ 
Mid-Late \textbf{+ MS + AL} 
& 83.8 {\tiny$\pm$1.1} & 0.769 {\tiny$\pm$0.016} & 73.0 {\tiny$\pm$1.5}\\
CoRe-Sleep
& 84.0 {\tiny$\pm$1.2} & 0.771 {\tiny$\pm$0.017} & 73.0 {\tiny$\pm$0.7}\\

\hdashline

XSleepNet \cite{phan2021xsleepnet}
& 75.5 \tiny$\pm$2.6 & 0.641 \tiny$\pm$0.04 & 61.7 \tiny$\pm$4.3 \\

\bottomrule
\end{tabular} 
\end{table}

\begin{figure*}[t]
    \begin{subfigure}[b]{0.33\textwidth}
        \centering
        \includegraphics[width=\textwidth]{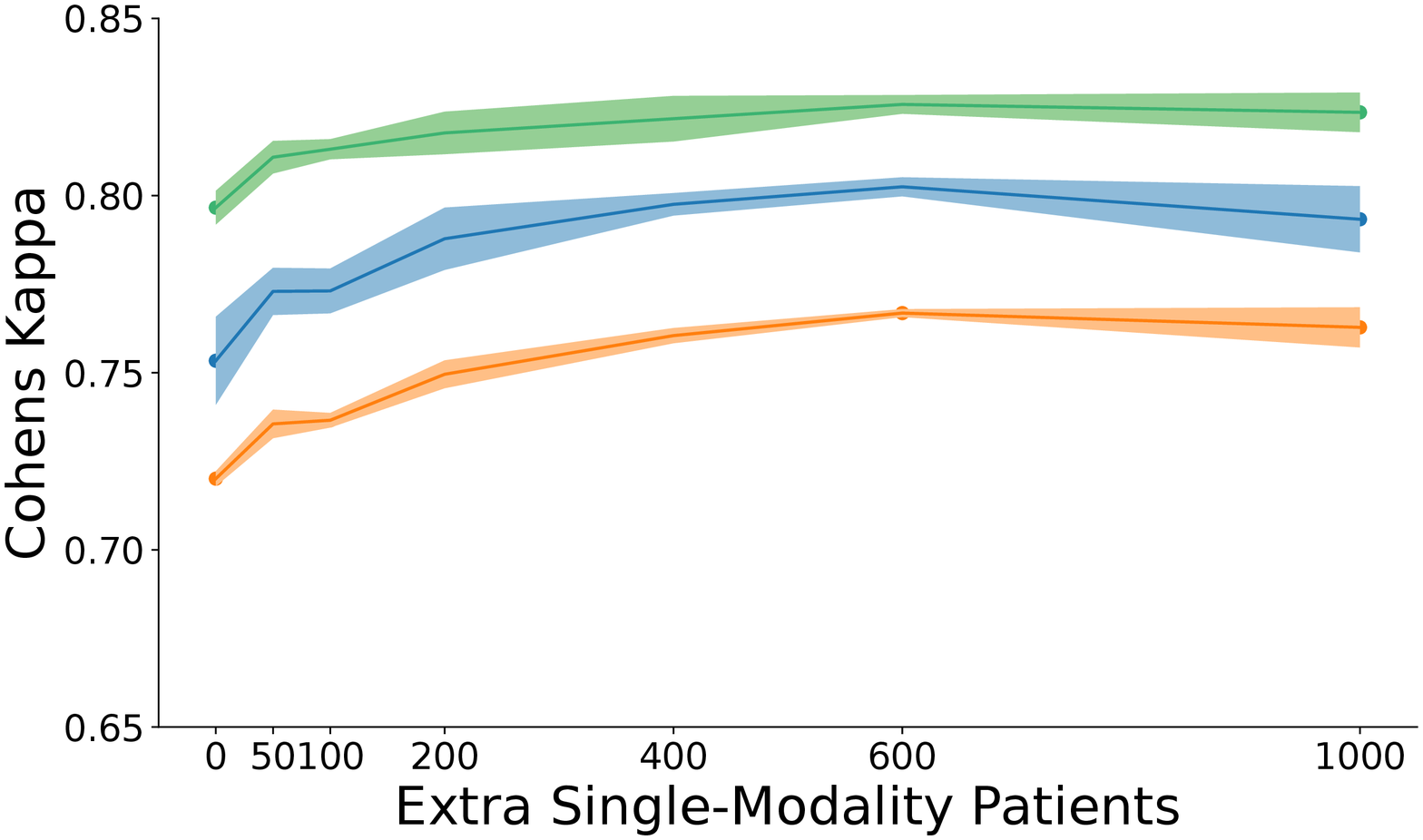}
        \caption{Unimodal EEG and EOG}
        \label{fig:incomplete_both}
    \end{subfigure}
    \begin{subfigure}[b]{0.33\textwidth}
        \centering
        \includegraphics[width=\textwidth]{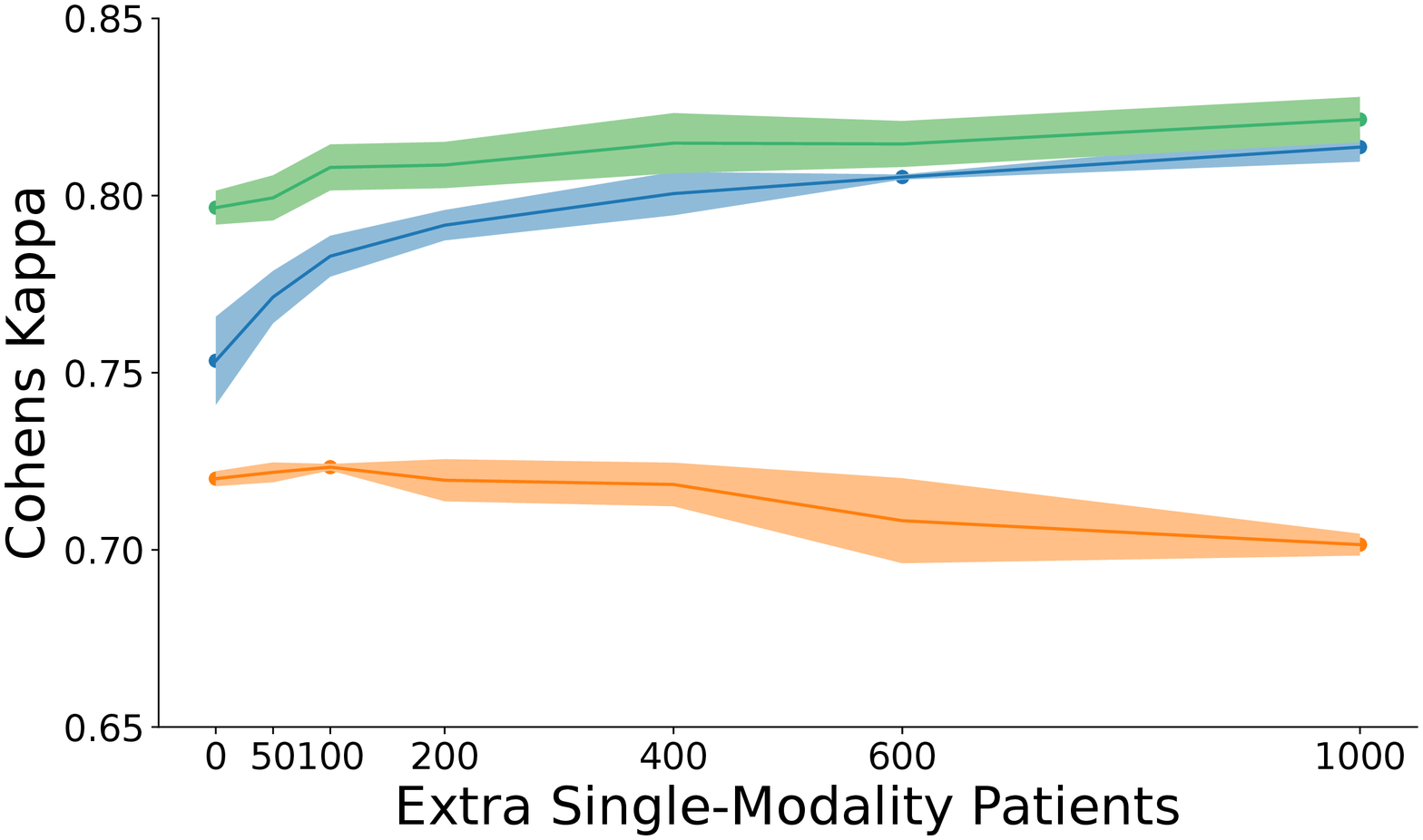}
        \caption{Unimodal EEG}
        \label{fig:incomplete_eeg}
    \end{subfigure}
  \begin{subfigure}[b]{0.33\textwidth}
    \centering
    \includegraphics[width=\textwidth]{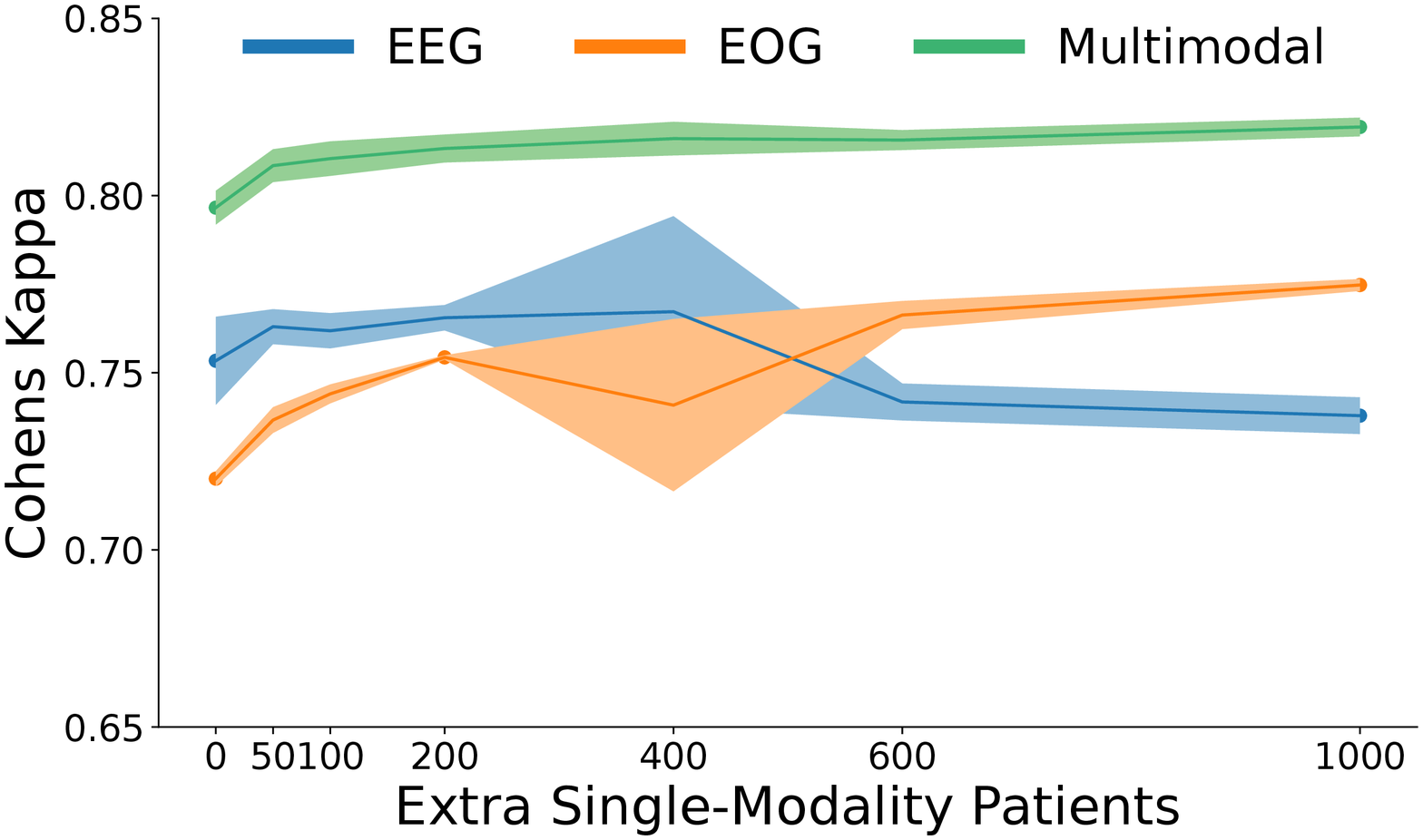}
    \caption{Unimodal EOG}
    \label{fig:incomplete_eog}
\end{subfigure}
  \caption{CoRe-Sleep was trained using a sub-group of 100 multimodal patients who had both EEG and EOG, as well as some unimodal patients who only had EEG or EOG: \textbf{a)} We add patients who provide only EEG, as well as an equal number of patients who provide only EOG. We consistently observe an improvement in the performance of both multimodal and unimodal predictors. \textbf{b)} We add patients who provide only EEG. A decrease of EOG predictor performance is noted. \textbf{c)} We add patients who provide only EOG. Consistently here, we notice a decrease in the performance of the modality that was not provided. All experiments have been repeated three times with different dataset splits to obtain the shown variance.}
\label{fig_incomplete_training}
\end{figure*}

\subsection{Noisy Modalities} 

A patient in this subset is visualised in Figure \ref{fig:time_std_preds_patient}. After almost two and a half hours, the EEG begins to behave irregularly, highly likely because the electrode is getting detached. The predictions of unimodal EEG in that area can be seen below in blue. When the recording is noisy, the predictions are clearly wrong as continuous Wake is predicted. The CoRe-Sleep model is less affected by noise in the EEG data, as it adapts its predictions based on information contained in the EOG. Evidently, the predictions of unimodal EOG and CoRe-Sleep in the noisy area display great similarities.  

In general, testing on the subset of noisy patients leads to decreased performance in all models. However, such imperfect data quality does occur in real life. Table \ref{table:noisy_shhs} shows an overview of the performance on the noisy subset. Since EEG is more often the broken modality in the subset, its unimodal equivalent model performs substantially worse than the EOG one. Models Early and CoRe underachieve in comparison to using EOG only. On the contrary, Mid-Late shows some further robustness to noisy cases. The Mid-Late model has separate encoders for each modality, allowing one to remain uninfluenced by the noisy behaviour of the other. It's the same property that allows it to do well in the missing modality scenario. Adding AL and MS losses benefits all three models. Only in the case of Mid-Late, we see some declined behaviour when both are added compared with adding only the MS loss. CoRe-Sleep manages to perform close to Mid-Late model exhibiting robustness in the noisy modality scenario. The added value of using multimodal fusion is highlighted since the predictions based on noisy modality can be amended by the non-noisy ones. Finally, the literature benchmark evaluated on multimodal data, XSleepNet, is an early fusion approach and clearly underperforms in this subset.

Evaluating a model solely on carefully screened data that exclude any noise or missing modalities, as done in most datasets, may not accurately reflect the full capabilities of the model. Testing the models on missing or noisy modalities can give a further less biased impression of the models' abilities.

\subsection{Training with Incomplete Data}

In the last experiment, the ability of CoRe-Sleep to be trained with the assistance of large sets of modality-incomplete data is examined. In Figure \ref{fig_incomplete_training}, three different training scenarios are presented. In each scenario, we train models with a different number of training patients. Every model has the same 100 multimodal patients from which both MS and AL loss are calculated. Furthermore, unimodal patients are added gradually, either from both modalities or from solely EEG or EOG in Figures \ref{fig:incomplete_both}, \ref{fig:incomplete_eeg} and \ref{fig:incomplete_eog} respectively. For those additional patients, we calculate only the unimodal loss with respect to the included modality. The generalization performance on the test set is measured for each step. In this experiment, parameter sharing among predictors facilitated faster and more stable convergence, particularly in situations where data was limited in size.

Three major findings can be observed:

\begin{enumerate}
  \item When the number of additional unimodal patients is close to the number of initial multimodal ones (in this case 100) then all three predictors increase their performance. This can be seen in all three experiments.

  \item Adding the same number of unimodal patients from both modalities, without being the same patients, favours all three predictors even when adding unimodal ones outmatch by far the initial multimodal ones (see Figure \ref{fig:incomplete_both}). 
    
  \item The performance of the other modality deteriorates when adding unimodal patients in much larger numbers than the initial multimodal ones, as shown in Figures \ref{fig:incomplete_eeg} and \ref{fig:incomplete_eog}.
  
\end{enumerate}

With this experiment, we show that patients' recordings with missing modalities can be included in the training and help to improve overall model performance. Adding unimodal data should be carefully exploited at a large scale (greater magnitudes than multimodal data), especially in cases where inference should come separately from both modalities. 
CoRe-Sleep can leverage all available data making it more flexible for real-world scenarios.

\section{Conclusion}
\label{sec:conclusion}

We propose CoRe-Sleep, a multimodal model that is able to perform inference on incomplete multimodal data and that shows robustness to missing modalities. CoRe-Sleep exceeds by a small margin previous state-of-the-art results in both multimodal and unimodal settings. Trained with a multi-task loss, it shows robustness in noisy modality scenarios. We compare against other architectures to reveal the properties of the Coordinate Representation fusion and the chosen multi-task loss. As noisy recordings cannot be avoided in real life, this work could be a significant step towards automating sleep staging in clinical settings.

Our study highlights two key insights regarding multimodal machine learning. Firstly, training models with multiple modalities can yield performance improvements, even when testing only involves a subset of the modalities. Contrary to common practice, training models on the specific modalities that will be present during testing may not always lead to optimal performance. Secondly, our findings suggest that including training data points that lack some of the modalities can benefit model performance. These insights can inform the development of more effective multimodal machine learning approaches for a wide range of applications.

\bibliographystyle{IEEEtran}
\bibliography{main}

\end{document}